# Controlling spin current polarization through non-collinear antiferromagnetism

T. Nan[1], C. X. Quintela[1], J. Irwin[2], G. Gurung[3], D. F. Shao[3], J. Gibbons[4], N. Campbell[2], K. Song[5], S. Y. Choi[5], L. Guo[1], R. D. Johnson[6,7,8], P. Manuel[8], R. V. Chopdekar[9], I. Hallsteinsen[9,10], T. Tybell[10], P. J. Ryan[11,12], J. W. Kim[11], Y. S. Choi[11], P. G. Radaelli[6], D. C. Ralph[4,13], E. Y. Tsymbal[3], M. S. Rzchowski[2], C. B. Eom[1*]

**Abstract**

**The spin-Hall effect describes the interconversion of charge currents and spin currents, enabling highly efficient manipulation of magnetization for spintronics[1–8]. Symmetry conditions generally restrict polarizations of these spin currents to be orthogonal to both the charge and spin flows[9,10]. Spin polarizations can deviate from such direction in nonmagnetic materials only when the crystalline symmetry is reduced[11]. Here we experimentally show control of the spin polarization direction by using a non-collinear antiferromagnet $Mn_3GaN$[12,13], in which the triangular spin structure creates a low magnetic symmetry state while maintaining a high crystalline symmetry. We demonstrate that epitaxial $Mn_3GaN$/Permalloy heterostructures can generate unique types of spin-Hall torques at room temperature corresponding to unconventional spin polarizations collinear to spin currents or charge currents which are forbidden in any sample with two-fold rotational symmetry. Our results demonstrate an approach based on spin-structure design for controlling spin-orbit torque, paving the way for further progress in the emergent field of antiferromagnetic spintronics[14–16].**

[1]Department of Materials Science and Engineering, University of Wisconsin-Madison, Madison, Wisconsin 53706, USA; [2]Department of Physics, University of Wisconsin-Madison, Madison, Wisconsin 53706, USA; [3]Department of Physics and Astronomy & Nebraska Center for Materials and Nanoscience, University of Nebraska, Lincoln, NE 68588, USA; [4]Laboratory of Atomic and Solid State Physics, Cornell University, Ithaca, New York 14853, USA; [5]Department of Materials Science and Engineering, POSTECH, Pohang 37673, South Korea; [6]Clarendon Laboratory, Department of Physics, University of Oxford, Parks Road, Oxford OX1 3PU, UK; [7]ISIS Facility, Rutherford Appleton Laboratory, Chilton, Didcot, OX11 0QX, UK; [8]Department of Physics and Astronomy, University College London, Gower Street, London, WC1E 6BT, UK; [9]Advanced Light Source, Lawrence Berkeley National Laboratory (LBNL), 1 Cyclotron Road, Berkeley, California 94720, USA; [10]Department of Electronic Systems, Norwegian University of Science and Technology, Trondheim 7491, Norway; [11]Advanced Photon Source, Argonne National Laboratory, Argonne, Illinois 60439, USA; [12]School of Physical Sciences, Dublin City University, Dublin 11, Ireland; [13]Kavli Institute at Cornell for Nanoscale Science, Ithaca, New York 14853, USA

* Corresponding author. Email: eom@engr.wisc.edu

In the classical picture of current-induced magnetization dynamics (Fig. 1f), charge currents in a multilayer sample flowing along the in-plane direction ($x$ direction) can generate out-of-plane spin currents (flowing in the $z$ direction) that have spin polarization $\sigma$, required by symmetry to be along the $y$ direction. This particular spin current can give rise to an anti-damping spin torque in an adjacent ferromagnet, which has magnetization vector $m$, of the form $m \times (m \times y)$. This anti-damping torque is responsible for efficient magnetization manipulation, but as it is restricted to lie along an in-plane direction, it is efficient for manipulating only samples with in-plane magnetic anisotropy. To efficiently and deterministically drive perpendicularly-magnetized devices that are preferred for high-density memories, an out-of-plane anti-damping torque is required. Such an unconventional spin-torque can originate from spin-orbit scattering from ferromagnetic interfaces[17–19], or can arise at the interface in systems with reduced symmetry, such as in bilayers of the non-magnetic transition-metal dichalcogenide and ferromagnetic metals[11]; however these effects based on interface or heterostructure engineering have not been demonstrated strong enough for practical anti-damping switching. Here we demonstrate an alternative strategy to achieve unconventional spin-orbit torques, based on long-range non-collinear magnetic order within the *bulk* of the spin-source layer. In particular, we use the spin-Hall effect in epitaxial thin films of Mn$_3$GaN, a metallic antiferromagnet that has a 120° triangular spin texture, which reduces the symmetry sufficiently to allow spin current generation with different spin polarization directions to generate unconventional spin-torques. When the non-collinear spin texture is eliminated by heating above the Néel temperature of Mn$_3$GaN (345 K), the unconventional spin-torques go to zero. Previously, although the spin-Hall effect has been demonstrated in antiferromagnetic thin films, only conventional spin-torque [$m \times (m \times y)$] has been observed[20–23].

Mn$_3$GaN is a metallic nitride with the antiperovskite crystal structure (identical to the perovskite structure, but with anion and cation positions interchanged) and a lattice parameter close to that of commonly used perovskite oxide substrates. In the bulk, it is known to exhibit antiferromagnetic ordering with a non-collinear $\Gamma^{5g}$ Kagome-like structure (magnetic space group: $R\bar{3}m$) stabilized by the magnetic frustration of the Mn atoms in the (111) plane (Fig. 1a)[12,24]. In the (001) plane of Mn$_3$GaN (Fig. 1b), the (110) plane is the only mirror plane. In this low-symmetry state, we find that charge currents along $x$ generate unconventional anti-damping torque components in the form of $\tau_x \propto m \times (m \times x)$ and $\tau_z \propto m \times (m \times z)$ in addition to the conventional $\tau_y \propto m \times (m \times y)$, which correspond to spin currents with $\sigma$ along $x$, $z$ and $y$, respectively (Fig. 1c). These spin-polarized currents have corresponding spin-Hall conductivities $\sigma_{zx}^x$, $\sigma_{zx}^z$ and $\sigma_{zx}^y$ (in the form of $\sigma_{jk}^i$, where $i$, $j$ and $k$ denote the spin polarization, spin flow and charge flow directions). The symmetry allowed, and experimentally observed, non-zero spin-Hall conductivities are consistent with our linear theory calculation (Supplementary Information). Figure 1d shows that the $\sigma_{zx}^x$, $\sigma_{zx}^z$ and $\sigma_{zx}^y$ calculated by using the bulk Mn$_3$GaN band structure are large within a wide energy window around the charge neutrality point, implying the existence of a sizable spin-Hall current even in the presence of charge carrier doping by defects. Above the antiferromagnetic-to-paramagnetic transition temperature (Néel temperature $T_N$), disordered spins give rise to a high-symmetry state (space group: $Pm\bar{3}m$) having 4 mirror planes in the crystal lattice (Fig. 1e), and consequently only the conventional spin-Hall conductivity $\sigma_{zx}^y$ can be non-zero. We list the matrices of the spin-Hall conductivity tensors, obtained from symmetry analysis and calculations, for Mn$_3$GaN in antiferromagnetic and paramagnetic phases in the Supplementary Information.

Epitaxial Mn$_3$GaN thin films were grown on (001) (La$_{0.3}$Sr$_{0.7}$)(Al$_{0.65}$Ta$_{0.35}$)O$_3$ (LSAT) substrates by reactive magnetron sputtering with *in-situ* reflection high-energy electron diffraction (RHEED, see Methods). The out-of-plane x-ray diffraction around the (002) LSAT substrate peak shows an epitaxial Mn$_3$GaN film (Fig. 2a). The distinct Kiessig fringes around the Mn$_3$GaN (002) peak (Fig. 2a) and the streaky RHEED pattern of the Mn$_3$GaN film surface (Fig. 2a inset) indicate a high crystalline quality and a smooth film surface. We also confirmed the cube-on-cube epitaxial relationship between the $Mn_3GaN$ film and underlying LSAT substrate (Supplementary Information). Ferromagnetic permalloy $Ni_{81}Fe_{19}$ (Py) and the Cu (as a spacer layer) thin films were then deposited *in situ* on $Mn_3GaN$ to form the Py/Cu/$Mn_3GaN$ bilayer, and finally were patterned into device bars for spin-torque measurements. In Fig. 2b, we show the cross-sectional filtered STEM-HAADF image of the bilayer, which reveals sharp interfaces between both $Mn_3GaN$/LSAT (left) and Py/$Mn_3GaN$ (right). Atomic force microscope images of the 10 nm Py/20 nm $Mn_3GaN$ surface indicate an atomically-smooth surface with a surface roughness of ~0.3 nm. Using neutron diffraction, , we determined that our 200-nm $Mn_3GaN$ films order with the bulk antiferromagnetic triangular $\Gamma^{5g}$ spin structure below a Néel temperature of $T_N$=350 K (Supplementary Information), slightly higher than for the thinner 20-nm films having $T_N$=345 K (see below). Using x-ray magnetic linear and circular dichroism with photoemission electron microsopy, we observe antiferromagnetic domains with size on the order of 200-300 nm (Supplementary Information). We note that domains with differing spin configurations can affect the unconventional spin torque, since the unconventional spin-Hall conductivity terms can be averaged out to zero under certain symmetry operations (Supplementary Information). The fact that we observe non-zero unconventional spin torques in $Mn_3GaN$, as described below, suggests that certain antiferromagnetic domain configurations are more favorable, which is inferred to be due to a tetragonal distortion that can induce a small non-compensated magnetic moment in $Mn_3GaN$ thin films. This unbalanced antiferromagnetic domain population is also evidenced by the finite x-ray magnetic linear dichroism (XMLD) signal from the $Mn_3GaN$ films at the Mn edge for a beam area 100's of microns in scale (Supplementary Information).

To measure the symmetry of the spin torques, we use the spin-torque ferromagnetic resonance (ST-FMR) technique (Fig. 3a)[11,25]. During the ST-FMR measurement, a microwave current applied to $Mn_3GaN$ produces alternating torques on the Py, and excites the Py magnetic moment into precession, generating a corresponding alternating sinusoidal change of the resistance $R$ due to the anisotropic magnetoresistance (AMR) of Py. We measure a dc voltage signal $V_{mix}$ across the device bar that arises from the mixing between the alternating current and changes in the device resistance. The resonance in $V_{mix}$ is obtained by sweeping the external in-plane magnetic field through the Py resonance condition (see Methods). Both in-plane and out-of-plane torque components can then be determined individually, as the symmetric and antisymmetric part of the line shape are proportional to the amplitude of the in-plane $\tau_\parallel$ and out-of-plane $\tau_\perp$ torque components, respectively. Considering only the conventional spin-Hall effect (or the Rashba-Edelstein effect and Oersted field), the in-plane and out-of-plane torque components would only have the form of $\boldsymbol{m} \times (\boldsymbol{m} \times \boldsymbol{y})$ and $\boldsymbol{m} \times \boldsymbol{y}$, respectively[8,26]. This corresponds to the case of samples containing materials with 2-fold rotational symmetry, in which case if $\boldsymbol{m}$ is inverted by rotating the in-plane magnetic field angle $\varphi$ (with respect to $\boldsymbol{x}$) by 180°, $V_{mix}$ must retain the same amplitude but change sign, giving $V_{mix}(\varphi) = -V_{mix}(\varphi + 180°)$. Any difference in the resonance line shape between $V_{mix}(\varphi)$ and $-V_{mix}(\varphi + 180°)$ indicates the presence of an additional, unconventional torque component.

Fig. 3b shows resonance spectra of the 10 nm Py/2 nm Cu/20 nm $Mn_3GaN$ sample with the current flow along the [100] direction for the magnetic field angle $\varphi$ equal to 40° and 220°, measured at room temperature when the $Mn_3GaN$ is in the antiferromagnetic state. The Cu insertion layer breaks the exchange coupling at the Py/$Mn_3GaN$ interface, but it allows the transmission of the spin current since Cu has a long spin diffusion length. We find that the $V_{mix}(40°)$ and $-V_{mix}(220°)$ scans are notably different in the antiferromagnetic phase, indicating the presence of unconventional torque components[11].

To examine the torque components quantitatively, we perform ST-FMR measurements as a function of the in-plane magnetic field angle $\varphi$. Fig. 3c and d shows the angular dependence of symmetric $V_S$ and antisymmetric $V_A$ part for the 10 nm Py/2 nm Cu/20 nm $Mn_3GaN$ sample, measured at room temperature.

The angular dependence of ST-FMR can be understood as the product of the AMR in Py [$dR/d\varphi \propto \sin(2\varphi)$], with the in-plane $\tau_\parallel$ or out-of-plane torque $\tau_\perp$ components, as $V_S \propto \sin(2\varphi)\tau_\parallel$ and $V_A \propto \sin(2\varphi)\tau_\perp$. For ferromagnetic metal/normal metal bilayers (i.e. Py/Pt), the conventional anti-damping torque $\boldsymbol{\tau_{y,AD}} \propto \boldsymbol{m} \times (\boldsymbol{m} \times \boldsymbol{y})$ and field-like torque $\boldsymbol{\tau_{y,FL}} \propto \boldsymbol{m} \times \boldsymbol{y}$ both have a $\cos(\varphi)$ dependence, giving rise to an overall angular dependence of the form $\sin(2\varphi)\cos(\varphi)$ for both $V_S$ and $V_A$. We find the angular dependence of both $V_S$ and $V_A$ for the Mn$_3$GaN clearly deviate from this simple model (Fig. 3c and d, grey line), but can be well fitted by adding additional, unconventional torque terms with the presence of spin currents with spin polarizations oriented away from $\boldsymbol{y}$. The spin currents that are polarized along $\boldsymbol{x}$ would generate torque [$\boldsymbol{\tau_{x,AD}} \propto \boldsymbol{m} \times (\boldsymbol{m} \times \boldsymbol{x})$ and $\boldsymbol{\tau_{x,FL}} \propto \boldsymbol{m} \times \boldsymbol{x}$] with a $\sin(\varphi)$ dependence; while the torques with spin polarization along $\boldsymbol{z}$ [$\boldsymbol{\tau_{z,AD}} \propto \boldsymbol{m} \times (\boldsymbol{m} \times \boldsymbol{z})$ and $\boldsymbol{\tau_{z,FL}} \propto \boldsymbol{m} \times \boldsymbol{z}$], since $\boldsymbol{m}$ is oriented in the plane, are independent of $\varphi$. We thus fit $V_{mix,S}(\varphi)$ and $V_{mix,A}(\varphi)$ to more general forms to take all possible torque terms into account:

$$V_{mix,S}(\varphi) = \sin(2\varphi)\left(\tau_{x,AD}\sin(\varphi) + \tau_{y,AD}\cos(\varphi) + \tau_{z,FL}\right) \quad (1)$$

$$V_{mix,A}(\varphi) = \sin(2\varphi)\left(\tau_{x,FL}\sin(\varphi) + \tau_{y,FL}\cos(\varphi) + \tau_{z,AD}\right). \quad (2)$$

From the fitting, we find non-zero anti-damping torque terms $\tau_{x,AD}$, $\tau_{y,AD}$ and $\tau_{z,AD}$ demonstrating the existence of unconventional torque originated from spin polarizations along $\boldsymbol{x}$ and $\boldsymbol{z}$. This is consistent with the symmetry-allowed spin currents derived from the non-collinear antiferromagnetic Mn$_3$GaN magnetic space group through the bulk spin-Hall effect. This mechanism is distinct from those previously reported in noncentrosymmetric systems, and in magnetic tri-layers[18,19]. The generation of the spin torque $\boldsymbol{\tau_{\nu,AD}}$ relative to the charge current density can be parameterized into the spin-torque ratio $\theta_\nu = \frac{\hbar}{2e}\frac{j_{s,\nu}}{j_c}$, where $j_{s,\nu}$ is the spin current density with the spin polarization along $\nu$ that is absorbed by the Py, and $j_c$ is the charge current density in Mn$_3$GaN estimated from a parallel-conduction model. We find at room temperature that $\theta_x = -0.013 \pm 0.0002$, $\theta_y = 0.025 \pm 0.0002$ and $\theta_z = 0.019 \pm 0.0005$. The out-of-plane field-like torque has the form $\boldsymbol{\tau_{y,FL}}$, dominated by the contribution from the current-induced Oersted field (see Methods) with no detectable $\tau_{x,FL}$ torque. In addition, we observe an in-plane field-like torque $\tau_{z,FL}$ with a large torque ratio of $\theta_{FL,z} = -0.15 \pm 0.0002$, which could be generated along with $\boldsymbol{\tau_{z,AD}}$ by the spin currents polarized along $\boldsymbol{z}$.

We further confirmed the correlation between the observed unconventional spin polarization and the non-collinear spin structures in Mn$_3$GaN by performing angular-dependent ST-FMR measurements across its antiferromagnetic-to-paramagnetic phase transition. The Néel temperature of the 20 nm Mn$_3$GaN thin film is determined to be 345 K by tracking the temperature dependence of the out-of-plane lattice parameter (Fig. 4a) because the magnetic phase transition of Mn$_3$GaN produces a region of strong negative thermal expansion[27]. Fig. 4b-d show the temperature dependence (300 K to 380 K) of the ratios between anti-damping torque components and the Oersted torque, $\tau_{y,AD}/\tau_{y,FL}$, $\tau_{x,AD}/\tau_{y,FL}$ and $\tau_{z,AD}/\tau_{y,FL}$ (extracted from the full angular dependent ST-FMR measured at each temperature, see Supplementary Information). The unconventional torque ratios $\tau_{x,AD}/\tau_{y,FL}$ and $\tau_{z,AD}/\tau_{y,FL}$ vanish when the sample temperature is above the Néel temperature, while the conventional component $\tau_{y,AD}/\tau_{y,FL}$ remains non-zero, with a weak peak near the transition temperature (a similar peak in $\tau_{y,AD}$ has been observed near the Curie temperature of Fe$_x$Pt$_{1-x}$ alloys[28]). The vanishing of $\tau_{x,AD}/\tau_{y,FL}$ and $\tau_{z,AD}/\tau_{y,FL}$ directly demonstrates the strong correlation between the non-collinear spin structure and the existence of the unconventional spin torques. We also find that the unconventional torques persist at temperatures well below the Néel temperature, but decrease gradually at lower temperature with the increase of the canted moment in Mn$_3$GaN (Supplementary Information).

In summary, we have demonstrated the generation of unconventional spin-orbit torque based on low-symmetry non-collinear spin ordering present in the bulk of an epitaxial antiferromagnetic thin film with an antiperovskite structure. Such unconventional torques can be robustly manipulated by controlling the antiferromagnetic ordering across the Néel temperature. This work provides essential insight into understanding how unconventional spin-orbit torques can arise in systems with lower crystalline or magnetic symmetry. In addition, our finding offers the possibility to design and control spin currents through manipulating the non-collinear spin order via strain, temperature, chemical doping, and possibly external excitation, opening new areas of research opportunities in antiferromagnetic spintronics[15,16,29].

**Methods**

**Sample growth, fabrication and characterization**

Epitaxial Mn$_3$GaN thin films were grown on (001)-oriented LSAT substrates by DC reactive magnetron sputtering using a stoichiometric Mn$_3$Ga target in a vacuum chamber with a base pressure of $1\times10^{-8}$ Torr. During the growth, the Mn$_3$GaN growth mode and surface crystalline structure were monitored by *in situ* reflection high energy electron diffraction (RHEED). The growth undergoes a 3D to 2D growth mode transition. The streaky RHEED pattern after the deposition implies a smooth film surface (Supplementary Information). The growth was performed at a substrate temperature of 550 °C and an Ar (62 sccm)/N$_2$ (8 sccm) atmosphere of 10 mTorr. After the Mn$_3$GaN growth, the sample was cooled down in vacuum. The Cu and Py thin films were subsequently sputter deposited at an Ar pressure of 3 mTorr. The atomically flat sample surface was verified using atomic force microscopy (Supplementary Information). We confirmed the thickness, epitaxial arrangement, and coherence of the Mn$_3$GaN films using x-ray reflectivity, x-ray diffraction, and reciprocal space mappings. The growth rate of Py and Cu films were calibrated using x-ray reflectivity.

We patterned the Py/Mn$_3$GaN sample by using photolithography followed by ion beam milling. Then 200 nm Pt/5 nm Ti electrodes were sputter deposited and defined by a lift-off procedure. Devices for ST-FMR were patterned into microstrips (20-50 μm wide and 40-100 μm long) with ground-signal-ground electrodes. Devices for electrical transport measurements were patterned into 100 μm wide and 500 μm long Hall bars.

**STEM measurements**

The STEM sample was prepared through mechanical polishing down to a thickness of ~10 μm by using the precise polishing system (EM TXP, Leica). The polished specimen was then ion-milled using a 1-3 keV Ar ion beam (PIPS II, Gatan) to make the hole for the STEM observation. Afterwards, a low energy milling was performed using 0.1 keV Ar beam to minimize the surface damage from the prior ion-milling process.

The atomic structures were observed using a STEM (JEM-ARM200F, JEOL) at 200 kV equipped with an aberration corrector (ASCOR, CEOS GmbH). The optimum size of the electron probe was ~ 0.8 Å. The collection semi-angles of the HAADF detector were adjusted from 68 to 280 mrad in order to collect large-angle elastic scattering electrons for clear Z-sensitive images. The obtained raw images were processed with a band-pass Wiener filter with a local window to reduce a background noise (HREM research Inc.).

**ST-FMR measurements**

During ST-FMR measurements, a microwave current at a fixed frequency was applied with the in-plane magnetic field swept from 0-0.2 T for driving the ferromagnetic layer Py through its resonance condition. The amplitude of the microwave current is modulated at a low frequency (1.713 kHz), and the mixing voltage is detected through a lock-in amplifier. For the low temperature ST-FMR measurements (including the room temperature results shown in Fig. 3), the device was wire bonded to a coplanar waveguide and then transferred into a liquid helium flow cryostat. For the high temperature measurements (Fig. 4), the sample is placed on a resistive heater with the device probed by the ground-signal-ground rf probe. The ST-FMR resonance line shape can be fitted to a sum of symmetric $V_S$ and antisymmetric $V_A$ Lorentzian components in the form $V_{mix} = V_{mix,S} \frac{W^2}{(\mu_0 H_{ext} - \mu_0 H_{FMR})^2 + W^2} + V_{mix,A} \frac{W(\mu_0 H_{ext} - \mu_0 H_{FMR})}{(\mu_0 H_{ext} - \mu_0 H_{FMR})^2 + W^2}$, where $W$ is the half-width-at-half-maximum resonance linewidth, $\mu_0$ is the permeability in vacuum and $H_{FMR}$ is the resonance field. The in-plane $\tau_\parallel$ and out-of-plane $\tau_\perp$ components are proportional to $V_{mix,S}$ and $V_{mix,A}$ components, which can be expressed as,

$$V_{mix,S} = -\frac{I_{rf}}{2}\left(\frac{dR}{d\varphi}\right)\frac{1}{\alpha(2\mu_0 H_{FMR} + \mu_0 M_{eff})}\tau_\parallel \quad (1)$$

$$V_{mix,A} = -\frac{I_{rf}}{2}\left(\frac{dR}{d\varphi}\right)\frac{\sqrt{1 + M_{eff}/H_{FMR}}}{\alpha(2\mu_0 H_{FMR} + \mu_0 M_{eff})}\tau_\perp \quad (2)$$

where $I_{rf}$ is the microwave current, $R$ is the device resistance as a function of in-plane magnetic field angle $\varphi$ due to the AMR of Py, $\alpha$ is the Gilbert damping coefficient, and $M_{eff}$ is the effective magnetization. The AMR of Py is determined by measuring the device resistance as a function of magnetic field angle with a field magnitude of 0.1 T. We calibrate the microwave current $I_{rf}$ by measuring the microwave current induced device resistance change due to Joule heating[30,31] (Supplementary Information). The in-plane and out-of-plane torques can be expressed as the angular dependence of the torque components with different spin polarization directions,

$$\tau_\parallel = \tau_{x,AD}\sin(\varphi) + \tau_{y,AD}\cos(\varphi) + \tau_{z,FL} \quad (3)$$
$$\tau_\perp = \tau_{x,FL}\sin(\varphi) + \tau_{y,FL}\cos(\varphi) + \tau_{z,AD} \quad (4).$$

The strength of the torque components can then be determined from equation(1-4) with the calibrated $I_{rf}$ values, from which we noticed that the primary contribution to $\boldsymbol{\tau_{y,FL}}$ is the current-induced Oersted field. The spin torque ratios can be expressed as,

$$\theta_x = \frac{\tau_{x,AD}}{\tau_{y,FL}}\frac{e\mu_0 M_s t_{Py} t_{MGN}}{\hbar}\sqrt{1 + M_{eff}/H_{FMR}} \quad (5)$$
$$\theta_y = \frac{\tau_{y,AD}}{\tau_{y,FL}}\frac{e\mu_0 M_s t_{Py} t_{MGN}}{\hbar}\sqrt{1 + M_{eff}/H_{FMR}} \quad (6)$$
$$\theta_z = \frac{\tau_{z,AD}}{\tau_{y,FL}}\frac{e\mu_0 M_s t_{Py} t_{MGN}}{\hbar} \quad (7),$$

where $M_s$ and $t_{Py}$ are the saturation magnetization and the thickness of Py; $t_{MGN}$ is the thickness of MGN. $\hbar$ is the reduced Planck's constant and $e$ is the electron charge. The saturation magnetization of Py was measured with SQUID magnetometry, and is indistinguishable from the effective magnetization determined by ST-FMR.

**Electrical transport measurements of Mn$_3$GaN**

Electrical transport measurements of Mn$_3$GaN films were performed directly on as-grown films wire-bonded in a four-corner van der Pauw geometry. Both sheet resistance and Hall resistance were measured as a function of temperature and magnetic induction in a Quantum Design Physical Property Measurement System. Film resistivity was computed by solving the van der Pauw equation in conjunction with film thickness as measured with x-ray reflectivity, while Hall resistance was calculated by summing two orthogonal Hall configurations. The longitudinal resistivity and the Hall resistance of Mn$_3$GaN films vs. temperature are reported in the Supplementary Information.

**Temperature dependence of neutron diffraction**

Single crystal neutron diffraction measurements were performed on the WISH time-of-flight diffractometer[32] at ISIS, the UK neutron and muon source. A stack of eight, approximately 250 nm thick (001) $Mn_3GaN$ film samples with lateral dimensions 10 x 8 mm, were co-aligned and oriented for the measurement of nuclear and magnetic diffraction intensities in the (HK0) reciprocal lattice plane. The sample was first mounted within a $^4$He cryostat, and diffraction patterns were collected from a base temperature of 1.5 K up to 300 K, in 25 K steps. The sample was transferred to a medium-range furnace, and diffraction patterns were then collected at 320, 340, 360 and 390 K.

**Temperature dependence of X-ray diffraction**

The x-ray diffraction data were acquired at beamline 6-ID-B at the Advanced Photon Source with 12 keV incident x-ray energy. The sample temperature was controlled employing an ARS high temperature cryostat. Data were collected with 5 K steps; and at each temperature the sample position was realigned with respect to the base-temperature reciprocal space matrix. The sample was mounted on a standard PSI Huber diffractometer. The representative temperature dependence of x-ray diffraction spectra around the LSAT (003) reflection can be found in Supplementary Information.

**Theoretical calculations**

The electronic band structure of $Mn_3GaN$ was calculated by using first-principles density functional theory (DFT) with Quantum ESPRESSO[33] and fully relativistic ultrasoft pseudopotentials[34]. The exchange and correlation effects were treated within the generalized gradient approximation (GGA)[35]. The plane-wave cut-off energy of 57 Ry and a 16 × 16 × 16 $k$-point mesh in the irreducible Brillouin zone were used in the calculations. Spin-orbit coupling and noncollinear $\Gamma^{5g}$ antiferromagnetism were included in all electronic structure calculations. The calculated band structures for $Mn_3GaN$ in antiferromagnetic and paramagnetic phases are shown in Supplementary Information.

The spin Hall effect is given by[36]

$$\sigma_{ij}^k = \frac{e^2}{\hbar} \int \frac{d^3\vec{k}}{(2\pi)^3} \sum_n f_{n\vec{k}} \Omega_{n,ij}^k(\vec{k}),$$

$$\Omega_{n,ij}^k(\vec{k}) = -2Im \sum_{n \neq n'} \frac{\langle n\vec{k}|J_i^k|n'\vec{k}\rangle \langle n'\vec{k}|v_j|n\vec{k}\rangle}{\left(E_{n\vec{k}} - E_{n'\vec{k}}\right)^2},$$

where $f_{n\vec{k}}$ is the Fermi-Dirac distribution for the $n$th band, $J_i^k = \frac{1}{2}\{v_i, s_k\}$ is the spin current operator with spin operator $s_k$, $v_j = \frac{1}{\hbar}\frac{\partial H}{\partial k_j}$ is the velocity operator, and $i, j, k = x, y, z$. $\Omega_{n,ij}^k(\vec{k})$ is referred to as the spin Berry curvature in analogy to the ordinary Berry curvature. In order to calculate the spin Hall conductivities, we construct the tight-binding Hamiltonians using PAOFLOW code[37] based on the projection of the pseudo-atomic orbitals (PAO)[38,39] from the non-self-consistent calculations with a 16 × 16 × 16 $k$-point mesh. The spin Hall conductivities were calculated using the tight-binding Hamiltonians with a 48 × 48 × 48 $k$-point mesh by the adaptive broadening method to get the converged values.

**Synchrotron spectroscopy and microscopy**

X-ray magnetic circular dichroism (XMCD) and X-ray magnetic linear dichroism (XMLD) spectroscopy were measured at beamline 4.0.2, and x-ray microscopy at beamline 11.0.1.1 (PEEM-3) at the Advanced Light Source (ALS). In spectroscopy, total-electron-yield mode was employed by monitoring the sample drain current, and a gazing incidence angle of 30° to the sample surface along the [110] direction to probe the magnetic state. Hysteresis curves were obtained by monitoring the Fe and Mn $L_3$-edge XMCD as function of applied field. The obtained dichroism energies giving information on the magnetic nature of the $Mn_3GaN$ were then used to image the domain texture by x-ray photoemission electron microscopy (XPEEM), also taken with x-rays at a 30° grazing incidence along the [110] direction. Images taken at the maximum dichroism energies as a function of polarization were normalized by pre-edge energy images in order to minimize any topographic and work function contrast while emphasizing the magnetic contrast of the $Mn_3GaN$ films.

**Acknowledgments**

This work was supported by the National Science Foundation under DMREF Grant DMR-1629270, AFOSR FA9550-15-1-0334 and the Army Research Office through Grant W911NF-17-1-0462. Transport and magnetization measurement at the University of Wisconsin–Madison was supported by the US Department of Energy (DOE), Office of Science, Office of Basic Energy Sciences under Award DE-FG02-06ER46327. Work at Cornell was supported by the National Science Foundation (DMR-1708499), and was performed in part at the Cornell NanoScale Facility, an NNCI member supported by NSF Grant NNCI-1542081 and also in part in the shared facilities of the Cornell Center for Materials Research which are supported by the NSF Materials Research Science and Engineering Centers program (DMR-1719875). This research used resources of the Advanced Photon Source, a U.S. Department of Energy (DOE) Office of Science User Facility operated for the DOE Office of Science by Argonne National Laboratory under Contract No. DE-AC02-06CH11357. This research used resources of the Advanced Light Source, which is a DOE Office of Science User Facility under contract no. DE-AC02-05CH11231.

**Author Contributions**

T.N. and C.B.E. conceived the research. C.B.E., M.S.R., E.Y.T., D.C.R, P.G.R., T.T. and S.Y.C. supervised the experiments. T.N., C.X.Q., and L.G. performed the sample growth and surface/structural characterizations. T.N. performed the device fabrication. T.N., and J.G performed ST-FMR measurements and analysis. N.C. and J.I. performed magnetic and electrical transport characterizations. G.G. and D.F.S. performed the theoretical calculations. K.S. and S.Y.C. carried out the high-resolution TEM experiments. R.D.J., P.M. and P.G.R. conceived and performed the neutron diffraction experiment. P.R., J.W.K. and Y.S.C. carried out the synchrotron diffraction measurements. R.V.C, and I.H. performed synchrotron spectroscopy measurements, and R.V.C. performed and analyzed synchrotron microscopy measurements. T.N., J.I., D.F.S., D.C.R and C.B.E. wrote the manuscript. All authors discussed the results and commented on the manuscript. C.B.E. directed the research.

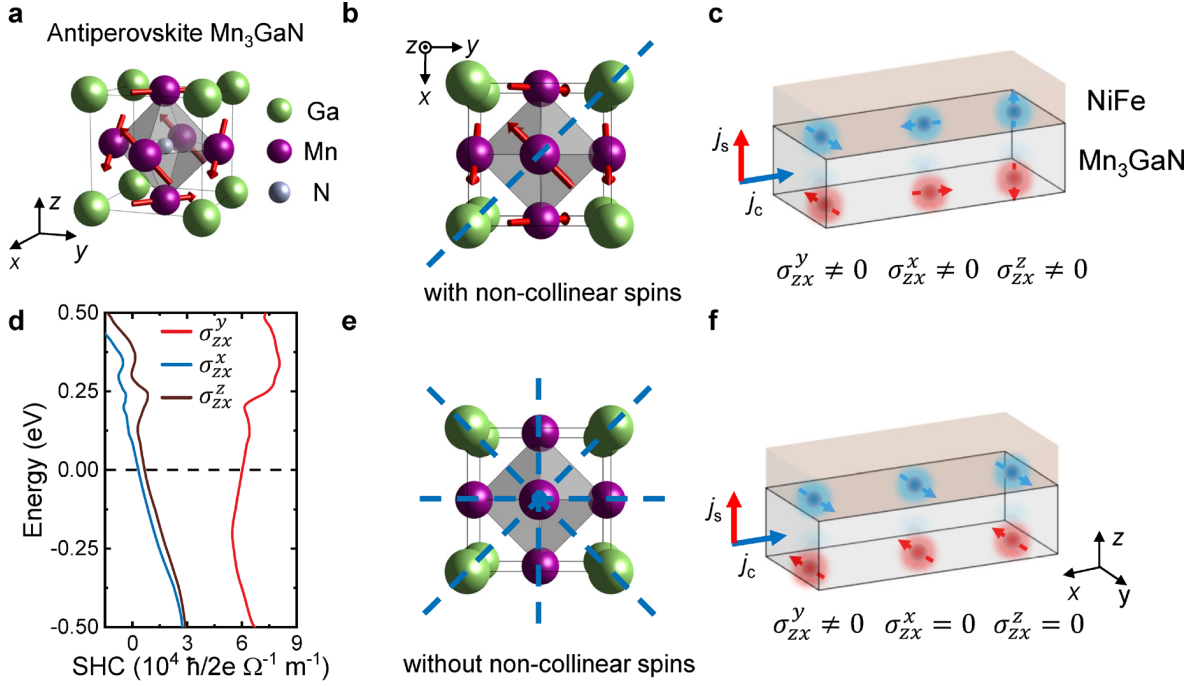

**Fig. 1 | The concept of the unconventional spin-Hall effect in Mn$_3$GaN. a**, The crystallographic unit cell of antiperovskite Mn$_3$GaN with the antiferromagnetic $\Gamma^{5g}$ spin structure where Mn spins (arrows) form a Kagome-type lattice in the (111) plane. *x*, *y*, and *z* correspond to the cubic [100], [010], and [001] axes, respectively. **b**, Spin structure of Mn$_3$GaN projected onto the (001) plane. The blue dashed line corresponds to the (110) mirror plane. **c**, Schematic illustrations of the Py/Mn$_3$GaN bilayer and the allowed spin-Hall spin polarization in the low-symmetry state (**a**). This indicates non-zero spin-Hall conductivities $\sigma_{zx}^y$, $\sigma_{zx}^x$ and $\sigma_{zx}^z$, which correspond to spin polarizations along *y*, *x* and *z* direction, respectively (with the charge current along *x* and spin flow along *z*). **d**, Calculated spin-Hall conductivities $\sigma_{zx}^y$, $\sigma_{zx}^x$ and $\sigma_{zx}^z$ for Mn$_3$GaN in the antiferromagnetic phase as a function of Fermi energy. **e**, Crystal structure of Mn$_3$GaN without non-collinear spin structure (i.e. above the antiferromagnetic transition temperature $T_N$) in the (001) plane, which gives rise to a high-symmetry state. **f**, Allowed spin polarization in the high-symmetry state, where only the conventional spin-Hall conductivity $\sigma_{zx}^y$ is non-zero.

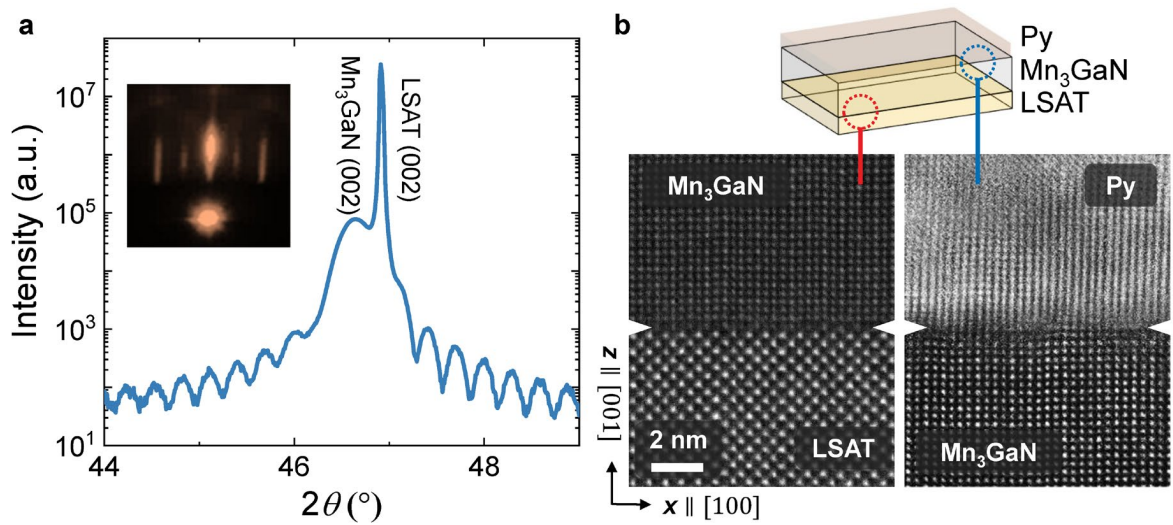

**Fig. 2 | Structural characterization of the Py/Mn₃GaN/LSAT. a**, 2θ-ω x-ray scan of the heterostructure of 10 nm Py/30 nm Mn₃GaN on LSAT (001) substrate showing single-phase Mn₃GaN with thickness oscillations indicating a smooth surface and sharp interface with the substrate. Inset shows reflection high-energy electron diffraction (RHEED) pattern of the specular diffraction spot for the Mn₃GaN surface. **b**, Scanning transmission electron microscope image of Py/Mn₃GaN heterostructure on (001) LSAT substrate with the top Py/Mn₃GaN interface (left), and the bottom Mn₃GaN/LSAT interface (right).

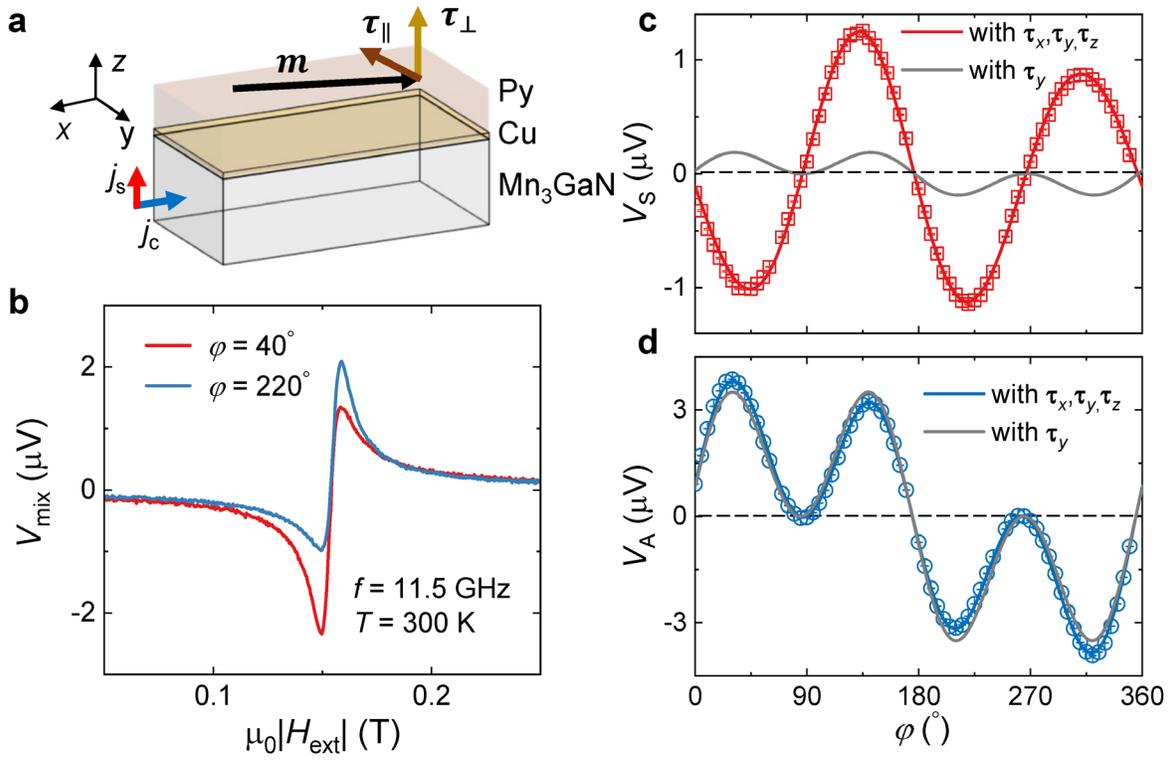

**Fig. 3 | ST-FMR measurements. a**, Schematic of the ST-FMR geometry for the Py/Cu/Mn$_3$GaN structures. $\tau_\parallel$ and $\tau_\perp$ denote the in-plane and out-of-plane torque components, which consist of different torque terms. **b**, ST-FMR spectra for the 10 nm Py/2nm Cu/20 nm Mn$_3$GaN device at 300 K (antiferromagnetic phase) with the Py magnetization oriented at 40° and 220° relative to the current axis. **c, d**, Symmetric (**c**) and antisymmetric (**b**) ST-FMR components for the 10 nm Py/2nm Cu/20 nm Mn$_3$GaN device as a function of the in-plane magnetic field angle at 300 K. The microwave current is applied along the [100] direction (*x* axis). The applied microwave frequency and power are 11.5 GHz and 15 dBm, respectively.

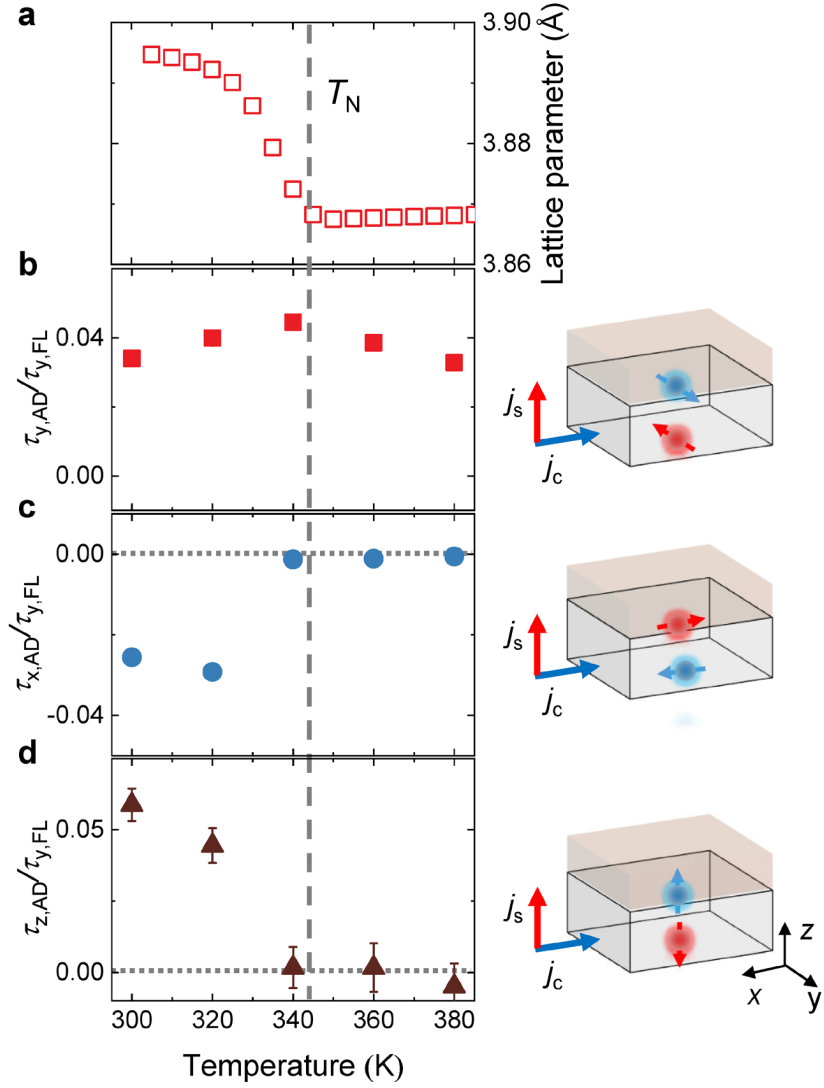

**Fig. 4 | Temperature dependence of spin-orbit torques. a,** Out-of-plane lattice parameter of a 30 nm Mn$_3$GaN/LSAT sample as a function of temperature; the lattice parameters anomaly indicates the Mn$_3$GaN Néel temperature $T_N$ of ~345 K. **b-d,** The torque ratios $\tau_{y,AD}/\tau_{y,FL}$, $\tau_{x,AD}/\tau_{y,FL}$ and $\tau_{z,AD}/\tau_{y,FL}$ as a function of the temperature. The schematics on the right panel show the geometry of the spin-Hall effect with different spin polarizations.

Supplementary Information for

**Controlling spin current polarization through non-collinear antiferromagnetism**

T. Nan[1], C. X. Quintela[1], J. Irwin[2], G. Gurung[3], D. F. Shao[3], J. Gibbons[4], N. Campbell[2], K. Song[5], S. Y. Choi[6], L. Guo[1], R. D. Johnson[7,8,9], P. Manuel[9], R. V. Chopdekar[10], I. Hallsteinsen[10,11], T. Tybell[11], P. J. Ryan[12,13], J. W. Kim[12], Y. S. Choi[12], P. G. Radaelli[7], D. C. Ralph[4,14], E. Y. Tsymbal[3], M. S. Rzchowski[2], C. B. Eom[1*]

[1]Department of Materials Science and Engineering, University of Wisconsin-Madison, Madison, Wisconsin 53706, USA; [2]Department of Physics, University of Wisconsin-Madison, Madison, Wisconsin 53706, USA; [3]Department of Physics and Astronomy & Nebraska Center for Materials and Nanoscience, University of Nebraska, Lincoln, NE 68588, USA; [4]Laboratory of Atomic and Solid State Physics, Cornell University, Ithaca, New York 14853, USA; [5] Department of Materials Modeling and Characterization, KIMS, Changwon 51508, South Korea; [6]Department of Materials Science and Engineering, POSTECH, Pohang 37673, South Korea; [7]Clarendon Laboratory, Department of Physics, University of Oxford, Parks Road, Oxford OX1 3PU, UK; [8]ISIS Facility, Rutherford Appleton Laboratory, Chilton, Didcot, OX11 0QX, UK; [9]Department of Physics and Astronomy, University College London, Gower Street, London, WC1E 6BT, UK; [10]Advanced Light Source, Lawrence Berkeley National Laboratory (LBNL), 1 Cyclotron Road, Berkeley, California 94720, USA; [11]Department of Electronic Systems, Norwegian University of Science and Technology, Trondheim 7491, Norway; [12]Advanced Photon Source, Argonne National Laboratory, Argonne, Illinois 60439, USA; [13]School of Physical Sciences, Dublin City University, Dublin 11, Ireland; [14]Kavli Institute at Cornell for Nanoscale Science, Ithaca, New York 14853, USA

* Corresponding author. Email: eom@engr.wisc.edu

**I. Structure and surface characterization of Py/Mn$_3$GaN/LSAT(001) heterostructures**

In Fig. 2a of the main text, we show the lab source out-of-plane x-ray diffraction of the Mn$_3$GaN (002) peak. In the 2$\theta$-$\omega$ scan with wide 2$\theta$ angles, we did not find any additional diffraction peak other than the (00$l$) reflections, indicating that the Mn$_3$GaN thin film is single phase, (00$l$) oriented. Fig. S1a shows the rocking curve of the Mn$_3$GaN (002) peak, where the full-width-half-maximum value is ~0.02° indicating the high crystalline quality of the film. The azimuthal $\Phi$-scan around the (022) reflection (Fig. S1b) indicates that the Mn$_3$GaN films grew epitaxially with a cube-on-cube alignment with the underlying LSAT substrate. The in-plane ($a$=3.916 Å) and out-of-plane ($c$=3.891 Å) lattice parameters of Mn$_3$GaN were determined from the x-ray reciprocal space mapping (RSM) measurements around the (-113) LSAT peak (bulk cubic Mn$_3$GaN lattice parameter of 3.898 Å), which gives rise to a tetragonal lattice. The tetragonal Mn$_3$GaN thin film breaks the cubic symmetry, potentially inducing a net magnetic moment[1], which is evidenced by the presence of an anomalous Hall effect near 100 K (Fig. S4b). Fig. S1c shows atomic force microscopy image of the final surface of the 10 nm Py/2nm Cu/20 nm Mn$_3$GaN/LSAT sample indicating an atomically flat surface with a RMS roughness <3 Å.

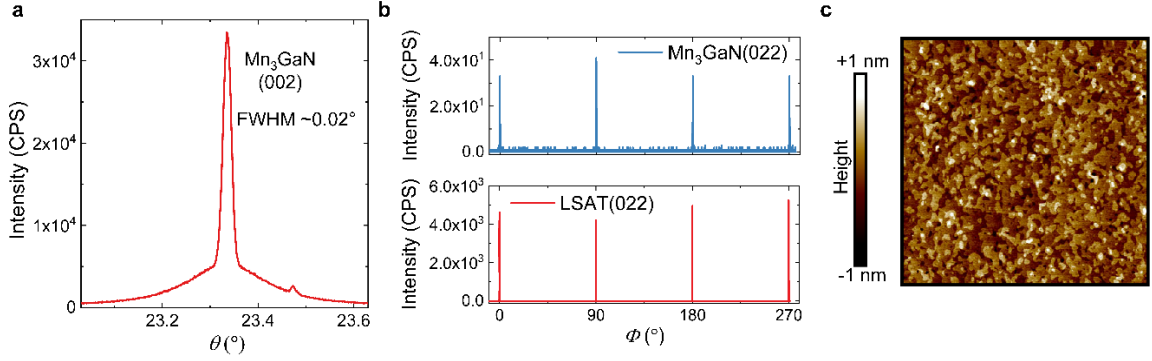

**Figure S1. a,** Rocking curve of the (002) Mn$_3$GaN peak. **b**, Φ-scan around the Mn$_3$GaN peak showing the epitaxial arrangement with the underlying LSAT substrate. **c**, Atomic force microscopy images of the heterostructure: 10 nm Py/20 nm Mn$_3$GaN on LSAT (001) substrate.

## II. Temperature dependence of x-ray diffraction

Bulk Mn$_3$GaN shows large negative thermal expansion behavior, which is linked to the first-order magnetic phase (Néel) transition. To identify the magneto-structural transition and thus determine the Néel temperature $T_N$ of our Mn$_3$GaN thin films, we performed x-ray diffraction experiments as a function of temperature. The temperature dependence of the out-of-plane lattice parameter was derived from the evolution of the Mn$_3$GaN (003) reflection, as shown in Fig. S2.

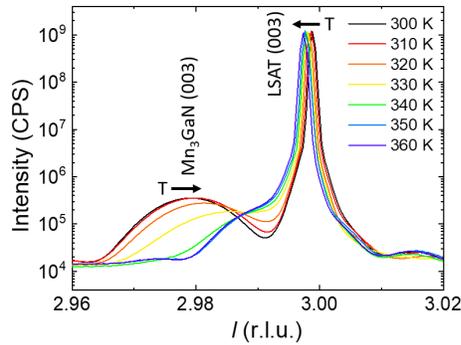

**Figure S2.** Temperature dependence of x-ray $l$ scans around the LSAT (003) reflection.

## III. Neutron diffraction of Mn$_3$GaN thin films

Single crystal neutron diffraction measurements were performed on a stack of eight, approximately 250 nm thick (001) Mn$_3$GaN film samples with lateral dimensions 10 x 8 mm, co-aligned and oriented for the measurement of nuclear and magnetic diffraction intensities in the (HK0) reciprocal lattice plane (see Methods section of the main text). Weak diffraction peaks were observed in the proximity of the strong (100) and (110) substrate Bragg peaks, as shown in Fig. S3 left. These weak peaks were found at positions corresponding to longer d-spacing values than the substrate reflections, consistent with the lattice mismatch between the Ma$_3$GaN film and substrate observed using x-ray diffraction. The temperature dependence of the integrated intensities of the weak peak is plotted in Fig. S3 right. Critical behavior was found for both intensities at $T_N$ ~340 K, which is consistent with the magnetic phase transition of the Ma$_3$GaN films observed in other measurements. Taken together, these observations allowed us to robustly assign the weak diffraction intensities to the Ma$_3$GaN film, with a magnetic component below $T_N$. Furthermore, calculations of the nuclear scattering intensity from Ma$_3$GaN confirmed that the (100) nuclear intensity is ~0, whilst the (110) is bright, as seen at 390 K.

The fact that the magnetic diffraction intensity coincided exactly with the nuclear intensity confirmed that the magnetic structure of the Ma$_3$GaN film below $T_N$ had a Γ-point propagation vector. Rigorous searches for diffraction peaks corresponding to other propagation vectors within the first Brillouin zone, as well as the monotonic behavior of the diffraction intensities, indicated that the Γ-point magnetic structure was the only one present at all measured temperatures below $T_N$, within the detection limit of the experiment.

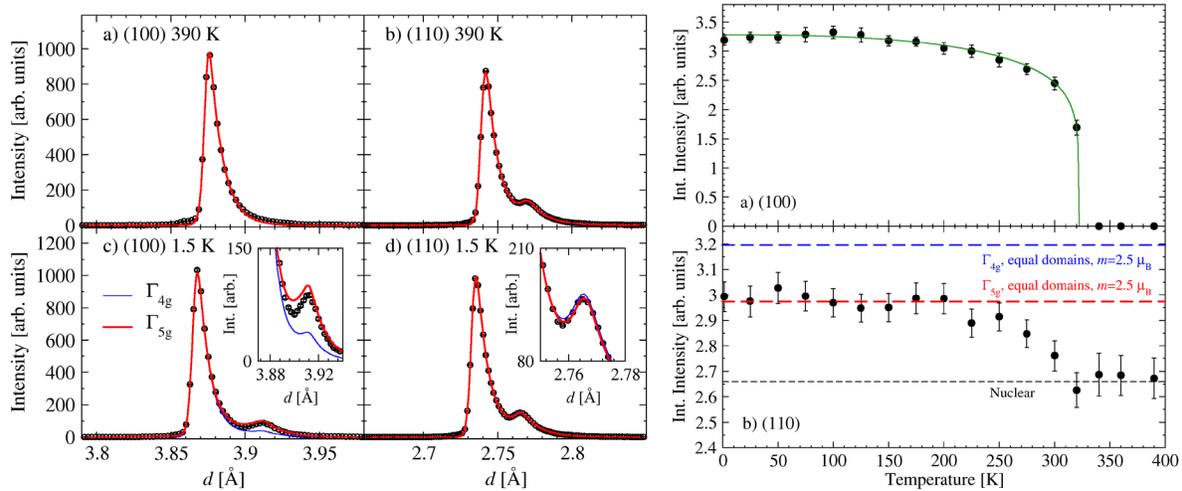

**Figure S3**. **Left**: Neutron diffraction data from a stack of 8 250-nm Mn$_3$GaN films on LSAT (001) substrates. The strong reflections on the left side are from the (100) **a**, **c** and (110) **b**, **d** Bragg peaks of the substrates. The side peak seen at 390 K in **b** is the (110) nuclear peak from the films. A side peak is absent in **a** because the (100) nuclear intensity from the film is ~0. Magnetic Bragg peaks at the Γ point (i.e., on top of the nuclear peaks of the film) develop at low temperatures **c**, **d**. Solid circles are experimental data, fitted with the Γ$_{4g}$ (blue lines) and Γ$_{5g}$ (red lines) representations. **Right**: The temperature dependence of the integrated intensity as a function of temperature for the (100) **a** and (100) **b** magnetic peaks. The dashed lines in **b** are the predictions of the magnetic intensity for the Γ$_{4g}$ (blue) and Γ$_{5g}$ (red) representations (see text). The green line in **a** is a guide to the eye.

Symmetry analyses using both the little group of the propagation vector, and the full Γ-point magnetic representation, were performed for the relevant Wyckoff positions using BasIreps (part of the FullProf package[2]) and Isodistort[34], respectively. Two three-dimensional irreducible representations appear in the decomposition of the full magnetic representation, typically labelled Γ$^{4g}$ and Γ$^{5g}$. Symmetry-distinct directions of the magnetic order parameter in the space spanned by both irreducible representations leads to 12 different magnetic symmetries. However, under the assumption that the magnetic structure has zero net magnetic moment (as evidenced by bulk magnetometry), and that every manganese site has the same moment magnitude, these 12 symmetries are reduced to just two with magnetic space groups $R$-$3m'$ and $R$-$3m$, which transform according to Γ$^{4g}$ and Γ$^{5g}$, respectively. In both symmetries, magnetic moments are aligned within the (111) crystallographic plane forming 120° triangular motifs. In the former case, moments lie within the mirror planes, and in the latter, perpendicular to the mirror planes. In fact, both magnetic structures are related by a 90° global rotation of spins in spin space, making them difficult to differentiate in diffraction, as discussed below.

The magnetic diffraction patterns for both $\Gamma^{4g}$ and $\Gamma^{5g}$ magnetic structures were calculated using FullProf[2] and used to fit the diffraction data (blue and red lines in Fig. S3 left, respectively). In both cases we assumed a Mn magnetic moment of 2.5 $\mu_B$ and equal population of all possible antiferromagnetic domains – a good assumption as the neutron diffraction experiment was performed using a stack of eight films, and the neutron beam illuminated the full volume of every film. It is clear that the $\Gamma_{5g}$ magnetic structure model is most consistent with the measured magnetic diffraction data, especially at the (100) reciprocal lattice point (Fig. S3 left). The sensitivity of the (110) reflection is most apparent when considering the relative magnitude of the magnetic intensity compared to the nuclear, as shown by the black, blue and red dashed lines in Fig. S3 right.

In summary, single crystal neutron diffraction experiments on our 250 nm thick $Mn_3GaN$ thin film stack demonstrated that long-range $\Gamma$-point ordering of manganese magnetic moments occurred below $T_N = 340$ K, and that the magnetic diffraction intensities at all measured temperatures below $T_N$ were fully consistent with the $\Gamma_{5g}$ magnetic structure previously proposed for bulk $Mn_3GaN$.

## IV. Electrical transport properties of $Mn_3GaN$ thin films

We measure as-grown 5 mm x 5 mm $Mn_3GaN$ thin films on LSAT in a Quantum Design PPMS in a van der Pauw geometry. Resistivity versus temperature data shown in Fig. S4a indicate that $Mn_3GaN$ is metallic. The slope of the Hall resistivity vs. field shows a sign change around 200 K. In semiconductors, such a sign change should be accompanied with a nonlinear Hall signal. However, as $Mn_3GaN$ is known to be highly covalent with many bands crossing the Fermi energy, none with particularly high mobility, the Hall curves are linear in field, and the sign change is caused by subtle band population changes with temperature. We also observed a small hysteresis in the Hall measurement around 100 K (red curve in Fig. S4b), indicating a presence of an anomalous Hall effect. The connection between ordinary Hall coefficient and magnetism allows us to confirm the Néel temperature at around 350 K by a flattening out of $R_H$ versus temperature (Fig. S4c).

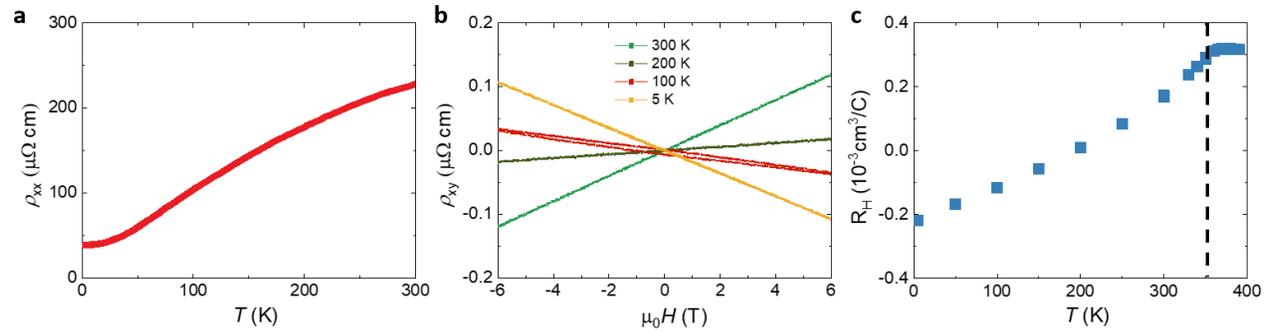

**Figure S4**. **a**, Longitudinal resistivity $\rho_{xx}$ vs. temperature curve for a 20 nm $Mn_3GaN$ thin film on LSAT (001) substrate, showing metallic behaviour. **b**, Hall resistivity $\rho_{xy}$ vs. out-of-plane magnetic field at various representative temperatures. **c**, Temperature dependence of ordinary Hall coefficient $R_H$, indicating a transition temperature at 350 K (dashed line).

## V. Magnetic properties of $Mn_3GaN$ thin films

We measure as-grown 5mm x 5mm $Mn_3GaN$ thin films on LSAT in a Quantum Design MPMS 3 with an in-plane applied field. The substrate contribution to the magnetization was measured separately and subtracted based on magnetic impurity density. Magnetization versus temperature data shown, in Fig. S5, have two clear transitions. The first is around 350 K, where the zero-field-cooled and field-cooled curves deviate. We ascribe this to the Néel Transition and note that the temperature matches with the flattening of the $R_H$ temperature dependence in Fig S4c. The second transition is around 200 K, and corresponds with a distinct further deviation of the field-cooled curve from the zero-field-cooled curve. This transition corresponds to the onset of anomalous Hall effect, shown partly in Fig S4b. Bulk work on $Mn_3GaN$ does not show any evidence for a new phase or net-moment character onset around 200 K, suggesting that this signal may not be due to an intrinsic or bulk mechanism.

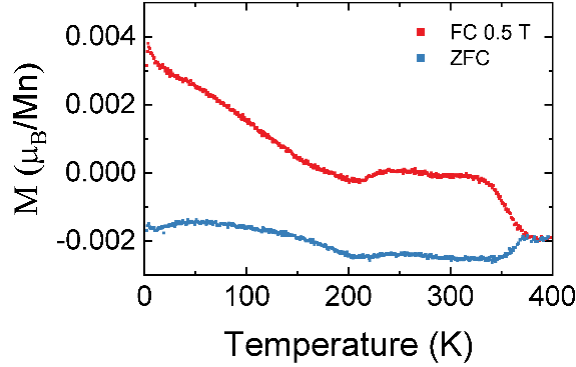

**Figure S5**. Net magnetization vs temperature curves for field-cooled (blue) and zero-field-cooled (red) samples. Distinct transitions are visible at 350K and 200K, and ascribed to the Neel temperature and the onset of the anomalous Hall effect respectively. The slight vertical offset comes from different ratios of magnetic impurities to intrinsic diamagnetism between the film substrate and reference substrate pieces.

## VI. ST-FMR line shape analysis

The ST-FMR signal with the current-induced in-plane and out-of-plane torque components is modeled by the Landau–Lifshitz–Gilbert–Slonczewski equation[5]. The ST-FMR mixing voltage can be then written in the form as,

$$V_{mix} = V_{mix,S} \frac{W^2}{(\mu_0 H_{ext} - \mu_0 H_{FMR})^2 + W^2} + V_{mix,A} \frac{W(\mu_0 H_{ext} - \mu_0 H_{FMR})}{(\mu_0 H_{ext} - \mu_0 H_{FMR})^2 + W^2} \quad [S1]$$

where $W$ is the half-width-at-half-maximum resonance linewidth, $\mu_0$ is the permeability in vacuum and $H_{FMR}$ is the resonance field. The symmetric $V_{mix,S}$ and the antisymmetric $V_{mix,A}$ Lorentzian amplitudes, which are proportional to the in-plane $\tau_\parallel$ and out-of-plane torque $\tau_\perp$ components, can be written as

$$V_{mix,S} = -\frac{I_{rf}}{2} \left(\frac{dR}{d\varphi}\right) \frac{1}{\alpha(2\mu_0 H_{FMR} + \mu_0 M_{eff})} \tau_\parallel \quad [S2]$$

$$V_{mix,A} = -\frac{I_{rf}}{2} \left(\frac{dR}{d\varphi}\right) \frac{\sqrt{1 + M_{eff}/H_{FMR}}}{\alpha(2\mu_0 H_{FMR} + \mu_0 M_{eff})} \tau_\perp, \quad [S3]$$

where $I_{rf}$ is the microwave current, $R$ is the device resistance as a function of in-plane magnetic field angle $\varphi$ due to the AMR of Py, $\alpha$ is the Gilbert damping coefficient, and $M_{eff}$ is the effective magnetization. The microwave current $I_{rf}$ with given microwave power is calibrated by measuring the device resistance change due to Joule heating effect[6,7]. We can compare the change of device resistance induced by the applied microwave power to that induced by the injection of a dc current $I_{dc}$. The rf current $I_{rf}$ can then be determined as $I_{rf} = \sqrt{2}I_{dc}$, since Joule heating from ac and dc current are $\frac{1}{2}I_{rf}^2 R$ and $I_{dc}^2 R$. Fig. S6a shows the resistance change for a typical device (10 nm Py/2nm Cu/20 nm Mn$_3$GaN/LSAT) as a function of dc current and rf power (at 7 GHz). To calibrate the anisotropic magnetoresistance $R(\varphi)$, we measure the device resistance as a function of magnetic field angle by rotating an in-plane magnetic field of 0.1 T produced by a rotary electromagnet. Fig. S6b shows the $dR/d\varphi$ as a function of magnetic field angle $\varphi$. The magnetic resonance properties were characterized by the frequency dependence of ST-FMR measurements. Fig. S6c shows the resonance linewidth $W$ as a function of frequency $f$. The Gilbert damping coefficient is calculated from $\alpha = \frac{|\gamma|}{2\pi f}(W - W_0)$, where $W_0$ is the inhomogeneous linewidth broadening, and $\gamma$ is the gyromagnetic ratio. From the linear fitting in Fig. S6c, we obtained $\alpha = 0.008$. The effective magnetization $M_{eff}$ is obtained by a fit of the resonance field $\mu_0 H_{FMR}$ as a function of frequency to the Kittel equation, $\mu_0 H_{FMR} = \frac{1}{2}\left[-\mu_0 M_{eff} + \sqrt{(\mu_0 M_{eff})^2 + 4\left(\frac{f}{\gamma}\right)^2}\right] - \mu_0 H_k$, where $\mu_0 H_k$ is the in-plane anisotropy field. As shown in Fig. S6d, the effective magnetization $M_{eff}$ is found to be 7.2×10$^5$ A/m.

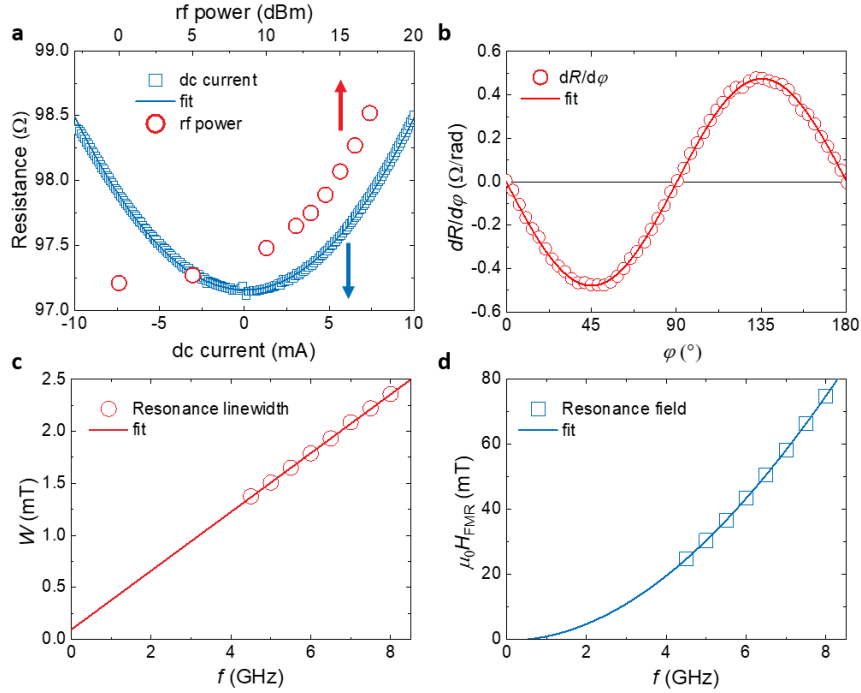

**Figure S6. a**, Resistance change as a function of applied dc current (blue) and microwave current (red), induced in a device (10 nm Py/2nm Cu/20 nm Mn$_3$GaN/LSAT) due to Joule heating. **b**, $dR/d\varphi$ vs. the magnetic field angle $\varphi$ derived from the anisotropic magnetoresistance of the same sample. **c,** The resonance linewidth $W$ as a function of frequency $f$. The solid curve shows the fit to a linear function, which gives a Gilbert damping coefficient of $\alpha =0.008$. **d**, Dependence of the resonance field $\mu_0 H_{FMR}$ upon frequency $f$. The data is fitted to the Kittel equation.

## VII. Angular dependent ST-FMR at different temperatures

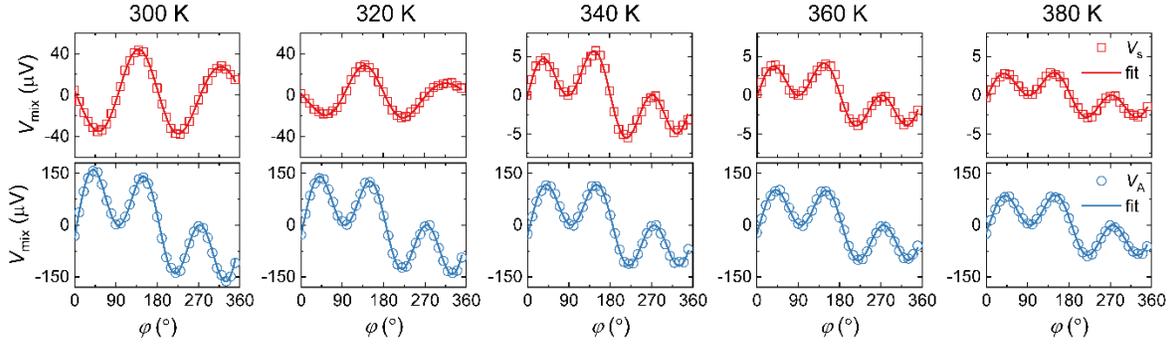

**Figure S7.** Angular dependence of the ST-FMR symmetric $V_{\text{mix,s}}$ (red) and the antisymmetric $V_{\text{mix,A}}$ (blue) components at different temperatures from 300 to 380 K, from which we extracted the temperature dependence of the torque ratios (shown in Fig. 4 of the main text).

## VIII. ST-FMR measurements at low temperatures

Fig. S8 shows the temperature dependence of the spin torque ratios $\theta_y$, $\theta_x$ and $\theta_z$ in the low temperature range 30-300 K. Interestingly, the conventional spin torque ratio $\theta_y$ (Fig. S8a) changes sign at ~200-250 K, which may correspond to the sign change of the Hall coefficient in Mn$_3$GaN at ~ 200 K (Fig. S4c). The amplitude of the unconventional torques $\theta_x$ (Fig. S8b) and $\theta_z$ (Fig. S8c) both decrease with decreasing temperature, which could be attributed to the increase of the canted moment in Mn$_3$GaN (Fig. S5).

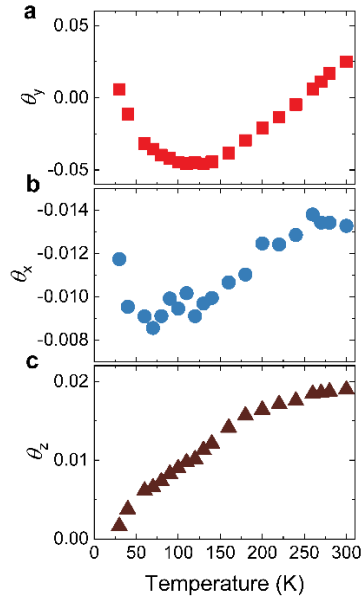

**Figure S8. a-c,** Spin torque ratios $\theta_y$, $\theta_x$ and $\theta_z$ as a function of temperatures.

## IX. Theory calculations

Fig. S9 shows the electronic band structure of $Mn_3GaN$ for different magnetic phases. In the paramagnetic phase, $Mn_3GaN$ has a space group of $Pm\bar{3}m$ (#221). When the $\Gamma^{5g}$ non-collinear antiferromagnetism is present below $T_N$, the symmetry of $Mn_3GaN$ is reduced to $R\bar{3}m$. The band structure of $\Gamma^{5g}$ phase is shown in Fig. S9b. The changes of symmetry due to magnetism significantly influence the electronic structure of $Mn_3GaN$, which also leads to the difference in spin-Hall conductivity between the $\Gamma^{5g}$ and the paramagnetic phases.

Table S1 summarizes the theoretical and calculated spin-Hall conductivity tensors derived from different $Mn_3GaN$ magnetic phases. As mentioned above and in the main text, paramagnetic $Mn_3GaN$ with the high-symmetry space group $Pm\bar{3}m$ only allows the conventional components in spin-Hall conductivity tensors[8]. The reduction of symmetry due to the non-collinear spin structure allows the unconventional components in spin-Hall conductivity tensors[8], which is consistent with our observation of unconventional spin-orbit torque.

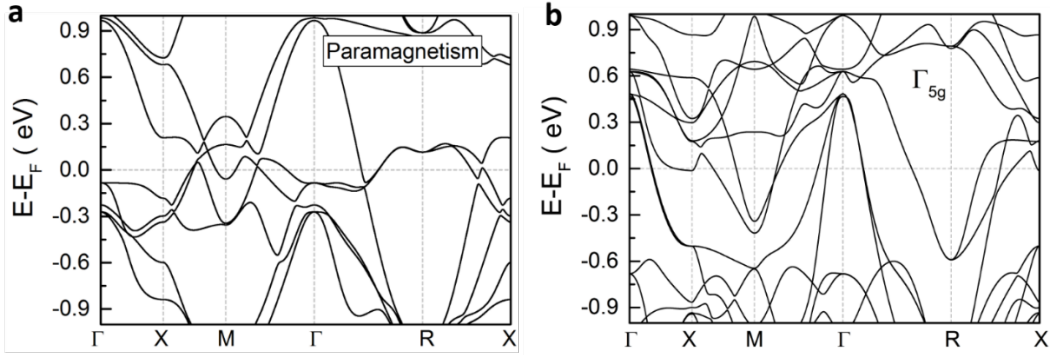

**Figure S9. a, b,** The electronic band structures of $Mn_3GaN$ in the paramagnetic (**a**) and in the antiferromagnetic $\Gamma^{5g}$ phases (**b**).

Table S1: The theoretical and calculated spin-Hall conductivity tensor for $\Gamma^{5g}$ phase and paramagnetic phase in $Mn_3GaN$.

|  |  | $\sigma^x$ | $\sigma^y$ | $\sigma^z$ |
|---|---|---|---|---|
| $\Gamma^{5g}$ | Tensor shape | $\begin{bmatrix} 0 & -\sigma_{yx}^y & \sigma_{yx}^y \\ \sigma_{yx}^x & -\sigma_{xx}^y & -\sigma_{yx}^z \\ -\sigma_{yx}^x & \sigma_{yx}^z & \sigma_{xx}^y \end{bmatrix}$ | $\begin{bmatrix} \sigma_{xx}^y & -\sigma_{yx}^x & \sigma_{yx}^z \\ \sigma_{yx}^y & 0 & -\sigma_{yx}^y \\ -\sigma_{yx}^z & \sigma_{yx}^x & -\sigma_{xx}^y \end{bmatrix}$ | $\begin{bmatrix} -\sigma_{xx}^y & -\sigma_{yx}^z & \sigma_{yx}^x \\ \sigma_{yx}^z & \sigma_{xx}^y & -\sigma_{yx}^x \\ -\sigma_{yx}^y & \sigma_{yx}^y & 0 \end{bmatrix}$ |
|  | Calculated tensor | $\begin{bmatrix} 0 & 45 & -45 \\ -29 & 40 & 114 \\ 29 & -114 & -40 \end{bmatrix}$ | $\begin{bmatrix} -40 & 29 & -114 \\ -45 & 0 & 45 \\ 114 & -29 & 40 \end{bmatrix}$ | $\begin{bmatrix} 40 & -114 & -29 \\ -114 & -40 & 29 \\ 45 & -45 & 0 \end{bmatrix}$ |

|              | Tensor shape    | $\begin{bmatrix} 0 & 0 & 0 \\ 0 & 0 & -\sigma_{yx}^z \\ 0 & \sigma_{yx}^z & 0 \end{bmatrix}$ | $\begin{bmatrix} 0 & 0 & \sigma_{yx}^z \\ 0 & 0 & 0 \\ -\sigma_{yx}^z & 0 & 0 \end{bmatrix}$ | $\begin{bmatrix} 0 & -\sigma_{yx}^z & 0 \\ \sigma_{yx}^z & 0 & 0 \\ 0 & 0 & 0 \end{bmatrix}$ |
|--------------|-----------------|---|---|---|
| Paramagnetic | Calculated tensor | $\begin{bmatrix} 0 & 0 & 0 \\ 0 & 0 & 41 \\ 0 & -41 & 0 \end{bmatrix}$ | $\begin{bmatrix} 0 & 0 & -41 \\ 0 & 0 & 0 \\ 41 & 0 & 0 \end{bmatrix}$ | $\begin{bmatrix} 0 & 41 & 0 \\ -41 & 0 & 0 \\ 0 & 0 & 0 \end{bmatrix}$ |

## X. Influence of antiferromagnetic domains

In magnets, symmetry generally requires the existence of degenerate domains, which strongly influence the properties of materials. For example, in noncollinear antiferromagnet $Mn_3Ge$ and $Ir_3Mn$, applying time reversal symmetry operation can reverse the moments and generate degenerate domains with opposite chirality, thus eliminating the anomalous Hall effect that is odd under time reversal symmetry[9]. Similar domains are also possible in our $Mn_3GaN$ film. The time reversal symmetry operation does not influence the spin-Hall conductivity. However, degenerate domains can be obtained by other symmetry operations. For example, a four-fold rotation symmetry around the $z$ direction leads to four degenerate domains, denoted as D1 to D4 in Fig. S10. The elements of the spin Hall conductivity tensors of $D_k$ (k=1,2,3,4) can be transformed from each other according to

$$\sigma_{ij,[D_k]}^S = \sum_{l,m,n} R_{z_{il}} R_{z_{jm}} R_{z_{kn}} \sigma_{lm,[D_n]}^S,$$

Where $R_{z_{il}}$ is an element of the rotation matrix $R_z$. Therefore, the average spin Hall conductivities can be obtained as

$$\bar{\sigma}_{ij}^S = \frac{1}{4}\sum_k \sigma_{ij,[D_k]}^S.$$

We found that if the four domains have the same fraction, the conventional components of the average spin Hall conductivities still exist, while the unconventional components will be cancelled. That deviates from our experimental observation, where the unconventional spin-Hall torques are robust.

We note, however, that a small tetragonal distortion exists in our sample. With the strain from the substrate, the tetragonality or *c/a* ratio for $Mn_3GaN$ is slightly smaller than 1. Such a tensile strain can introduce a small in-plane net magnetic moment along [110] directions[1]. The presence of such an in-plane net moment ensures the control of the magnetic order parameters in the domains[10,11]. We note that the $\Gamma^{5g}$ representation, which was robustly established by neutron diffraction for the 250 nm samples, does not admit a FM moment or anomalous Hall effect. The fact that both seem to exist in the thinner 20-nm samples emphasize the possible role of strain in modifying the magnetic and magnetotransport properties of this material.

Table S2: The spin Hall conductivity tensors for different domains.

| | $\sigma^x$ | $\sigma^y$ | $\sigma^z$ |
|---|---|---|---|
| General tensor | $\begin{bmatrix} \sigma_{xx}^x & \sigma_{xy}^x & \sigma_{xz}^x \\ \sigma_{yx}^x & \sigma_{yy}^x & \sigma_{yz}^x \\ \sigma_{zx}^x & \sigma_{zy}^x & \sigma_{zz}^x \end{bmatrix}$ | $\begin{bmatrix} \sigma_{xx}^y & \sigma_{xy}^y & \sigma_{xz}^y \\ \sigma_{yx}^y & \sigma_{yy}^y & \sigma_{yz}^y \\ \sigma_{zx}^y & \sigma_{zy}^y & \sigma_{zz}^y \end{bmatrix}$ | $\begin{bmatrix} \sigma_{xx}^z & \sigma_{xy}^z & \sigma_{xz}^z \\ \sigma_{yx}^z & \sigma_{yy}^z & \sigma_{yz}^z \\ \sigma_{zx}^z & \sigma_{zy}^z & \sigma_{zz}^z \end{bmatrix}$ |
| D1 | $\begin{bmatrix} 0 & -a & a \\ b & -c & -d \\ -b & d & c \end{bmatrix}$ | $\begin{bmatrix} c & -b & d \\ a & 0 & -a \\ -d & b & -c \end{bmatrix}$ | $\begin{bmatrix} -c & -d & b \\ d & c & -b \\ -a & a & 0 \end{bmatrix}$ |
| D2 | $\begin{bmatrix} 0 & a & -a \\ -b & -c & -d \\ b & d & c \end{bmatrix}$ | $\begin{bmatrix} -c & -b & d \\ a & 0 & a \\ -d & -b & c \end{bmatrix}$ | $\begin{bmatrix} c & -d & b \\ d & -c & b \\ a & -a & 0 \end{bmatrix}$ |
| D3 | $\begin{bmatrix} 0 & a & a \\ -b & c & -d \\ -b & d & -c \end{bmatrix}$ | $\begin{bmatrix} -c & b & d \\ -a & 0 & -a \\ -d & b & c \end{bmatrix}$ | $\begin{bmatrix} -c & -d & -b \\ d & c & b \\ a & -a & 0 \end{bmatrix}$ |
| D4 | $\begin{bmatrix} 0 & -a & -a \\ b & -c & -d \\ b & d & -c \end{bmatrix}$ | $\begin{bmatrix} c & b & d \\ -a & 0 & a \\ -d & -b & -c \end{bmatrix}$ | $\begin{bmatrix} c & -d & -b \\ d & -c & -b \\ a & a & 0 \end{bmatrix}$ |
| Average | $\begin{bmatrix} 0 & 0 & 0 \\ 0 & 0 & -d \\ 0 & d & 0 \end{bmatrix}$ | $\begin{bmatrix} 0 & 0 & d \\ 0 & 0 & 0 \\ -d & 0 & 0 \end{bmatrix}$ | $\begin{bmatrix} 0 & -d & 0 \\ d & 0 & 0 \\ 0 & 0 & 0 \end{bmatrix}$ |

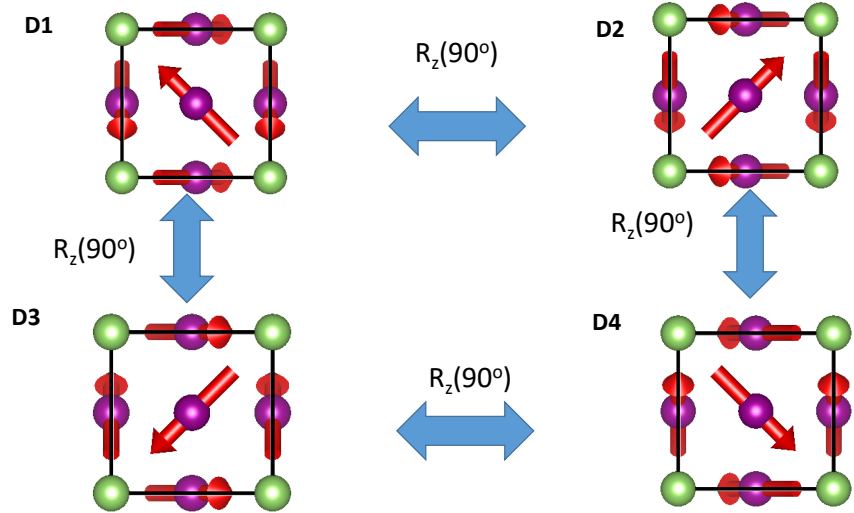

**Figure S10.** Four domains of Mn$_3$GaN generated by a four-fold rotation around $z$ direction.

## XI. Synchrotron spectroscopy and microscopy on Mn$_3$GaN films

In order to probe the antiferromagnetic domain structure we performed a combination of soft X-ray absorption spectroscopy and microscopy. Figure S11 depicts room temperature X-ray absorption spectroscopy (XAS), X-ray magnetic circular dichroism (XMCD) and X-ray magnetic linear dichroism (XMLD) measurements on a 4 nm Py/ 35 nm Mn$_3$GaN sample on the LSAT substrate (without a Cu spacer). XMCD spectroscopy measurements, in a grazing incidence geometry with a 0.3 T magnetic field applied along the [110] direction, indicate that a net ferromagnetic moment is present in both the permalloy (green curve in Fig. S11a) and Mn$_3$GaN (blue curve in Fig. S11b) layers.

X-ray magnetic linear dichroism (XMLD) at the Mn $L_{2,3}$ edge was measured in a normal-incidence x-ray geometry with magnetic field held along one of the [110] type in-plane directions and x-ray polarization axis projected along both orthogonal [110] type axes. The non-zero XMLD spectral intensity confirms that the population of canted antiferromagnetic domains can be influenced by the rotation of the Py layer and is not simply a random and equal distribution of domain variants as elaborated on in section X. We note that if the antiferromagnetic domain population were determined by the bulk domain variant degeneracy, the linear dichroism projections averaged across the beam area (approximately 150 x 500 microns) would cancel each other out and no net XMLD intensity would be measured. Furthermore, no evidence of oxidation was seen at the Fe (black curve left panel) or Mn (black curve middle panel) $L_{2,3}$ edge XAS lineshape. Hence the XAS data indicate that metal oxide antiferromagnetic phases are not present and that the XMLD signal hence originates from the Mn$_3$GaN layer.

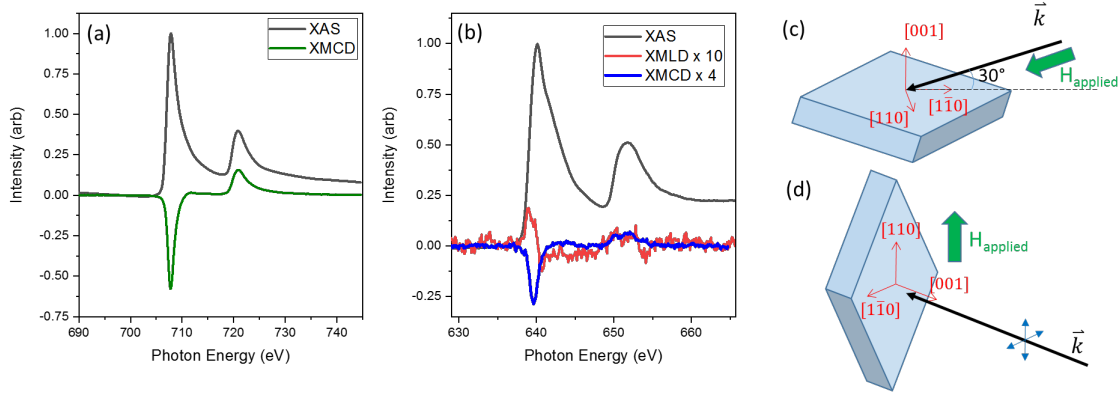

**Figure S11.** Room temperature magnetic spectroscopy on MGN/Permalloy bilayer at the (a) Fe and (b) Mn L edges. Field-dependent XMLD indicates an unequal population of antiferromagnetic domains, and XMCD measurements indicate that net ferromagnetic moments are present at the Mn edge. The measurement geometry was (c) grazing incidence for the PEEM and XMCD spectroscopy measurements and normal incidence for the XMLD data. A 0.3 T magnetic field was applied along the beam direction for XMCD spectroscopy and slightly canted towards [001] from the [110] direction for XMLD spectroscopy.

To further probe the non-equal domain population, spatially resolved XMLD-PEEM and XMCD-PEEM mappings of a single layer Mn$_3$GaN sample (without the Py capping) were taken in zero magnetic field at the peak energies in dichroism (from Fig. S11). Fig. S12 shows the XMLD-PEEM and XMCD-PEEM images obtained with the x-ray incidence direction along an in-plane [110] direction of the Mn$_3$GaN, revealing micron-sized XMLD domains and sub-micron sized XMCD domain regions at room temperature. There is a correlation between the contrast of the XMLD and XMCD domain locations; within each bright XMLD region there is strong contrast between XMCD domains, while in dark XMLD regions, there is much weaker XMCD contrast. This suggests that XMCD can be used to identify sub-regions of the XMLD regions, with the brightest XMCD domains having a small net moment along [1$\bar{1}$0] and darkest XMCD domains having a net moment along [$\bar{1}$10] according to the schematic in Fig. S11(c). The characteristic length scale of these canted antiferromagnetic domains is on the order of 200-300 nm. We note that the correlation between XMCD and XMLD domains observed here is consistent with the possible antiferromagnetic domain variants as described in Fig. S10.

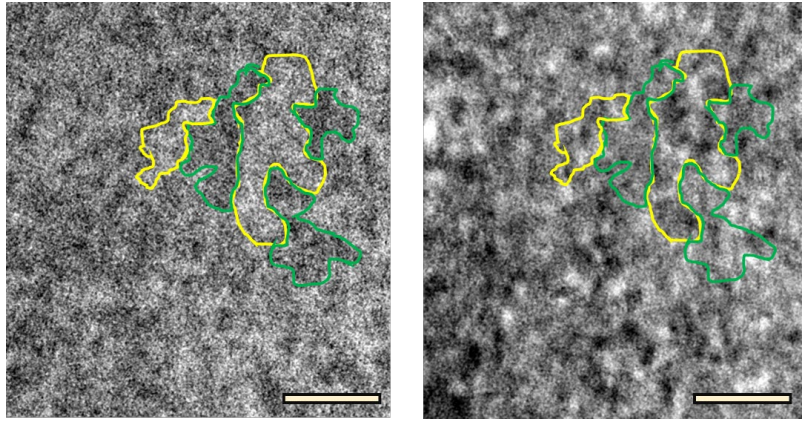

**Figure S12.** Comparison of Mn L edge XMLD (left) and XMCD (right) images at room temperature for a single layer Mn$_3$GaN film in the same region of the film. The x-rays are incident along the in-plane [110] direction and the scale bars are 1 micron.

To verify that the XMCD domain contrast is due to the frustrated antiferromagnetic order in Mn$_3$GaN, the sample was heated to above the Néel temperature, and then cooled to room temperature and imaged in the same region. Fig. S13 illustrates that the XMCD domain contrast vanishes at 360 K, and a different pattern of XMCD domains emerges after the sample is cooled to room temperature, albeit at a lower contrast level. This shows that the XMCD contrast is not localized to specific structural defects at the surface of the film but behaves similarly to other ferromagnetic and antiferromagnetic systems when cycled above their transition temperature.

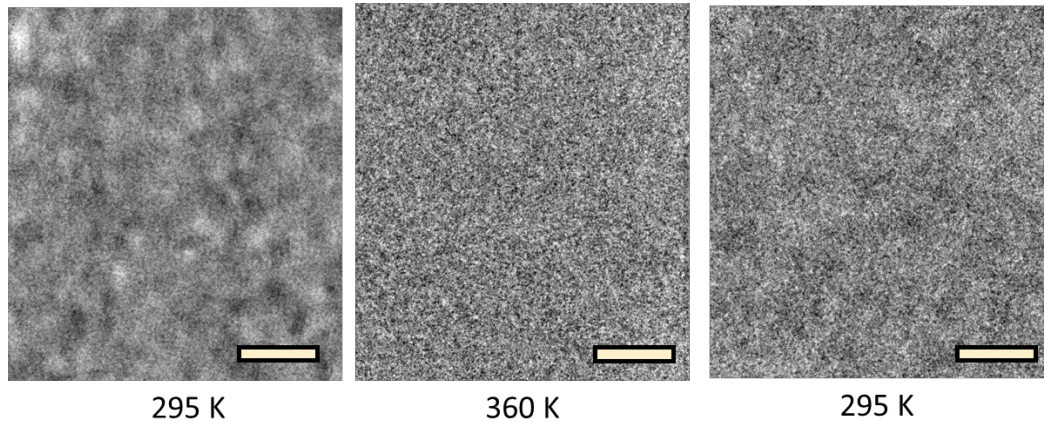

295 K       360 K       295 K

**Figure S13.** Mn L-edge XMCD images as a function of temperature for a single layer Mn$_3$GaN film taken in the same region of the sample with the same field of view. The x-rays are incident along the in-plane [$\bar{1}$10] direction and the scale bars are 1 micron. Contrast levels set to the same range (+/- 0.5% XMCD asymmetry) for all images. Domain contrast disappears by 360 K (center), and randomized domains re-emerge when the sample is cooled back down to room temperature (right).